\documentclass[aps,prl,twocolumn,groupedaddress,showpacs,amsmath,amssymb]{revtex4}

\newcommand\thintilde{{\lower.74ex\hbox{\mathtt{\char`\~}}}}
\usepackage{graphics}
\usepackage{color}
\usepackage{graphicx}
\usepackage{natbib}
\def \figurewidth {7.5cm}

\begin{document}

\title{Subwavelength position sensing using nonlinear feedback and wave chaos}

\author{Seth D. Cohen, Hugo L. D. de S. Cavalcante and Daniel J. Gauthier}

\affiliation{Department of Physics, Duke University, Department of Physics, Durham, North Carolina 27708, USA}

\date{\today}

\begin{abstract}
We demonstrate a position-sensing technique that relies on the inherent sensitivity of chaos, where we illuminate a subwavelength object with a complex structured radio-frequency field generated using wave chaos and a nonlinear feedback loop. We operate the system in a quasi-periodic state and analyze changes in the frequency content of the scalar voltage signal in the feedback loop. This allows us to extract the object's position with a one-dimensional resolution of $\sim\lambda/$10,000 and a two-dimensional resolution of $\sim\lambda/300$, where $\lambda$ is the shortest wavelength of the illuminating source. 
\end{abstract}

\pacs{05.45.Gg, 05.45.Mt, 84.30.Ng}
\maketitle

Diffraction, a property of electromagnetic (EM) waves, blurs spatial information less than the wavelength $\lambda$ of an illuminating source and hence limits the resolution of images. Over the past decade, techniques have been developed that overcome this diffraction limit using super-lenses made from negative-index media \cite{Zhang2008,Zhu2011}, super-oscillations \cite{Huang2007}, and nano-structures with surface plasmons \cite{Barnes2003,Anker2008}. Other methods use fluorescent molecules that serve as subwavelength point markers \cite{Rittweger2009,Zhuang2009,Gustafsson2005,Heintzmann2009}, where imaging is enabled by sensing the position of the markers. 

In this Letter, we describe a new super-resolution technique that senses the position of an object by combining two concepts: nonlinear delayed feedback and wave chaos. The system uses radio frequency (RF) EM waves in a closed feedback-loop through a wave-chaotic cavity. Self oscillation in the feedback occurs when the loop gain exceeds the loop losses; no RF field is supplied by an external source. We include a nonlinear element (NLE) in the feedback loop to create a system with complex (non-periodic) dynamics. The resulting EM oscillations provide the illumination source for our position sensor. 

Our work extends the Larsen effect, a known phenomenon where positive audio-feedback between a microphone and audio amplifier results in periodic acoustic oscillations. The frequency of oscillation, known as the Larsen frequency, is highly dependent on the propagation paths of the acoustic wave. A perturbation to these propagation paths shifts the Larsen frequency \cite{Lobkis2009}. For our super-resolution position-sensing system, we exploit the sensitivity of quasi-periodic EM frequencies.

\begin{figure}[thb]
\begin{center}
 \resizebox{8.0cm}{!}{\includegraphics{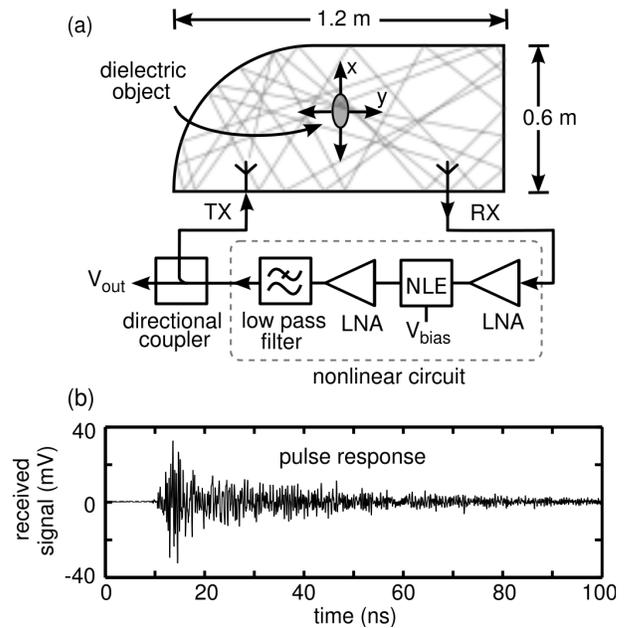}}
\end{center}
\caption{\label{fig:Figure1}(a) Experimental setup. Nonlinear circuit consisting of a transistor-based (BFG620) NLE, low-noise amplifiers (LNA, Mini-Circuits ZX60-4016E and Picosecond Pulse Labs 5828-108), and a low-pass filter. A bias voltage $V_{\text{bias}}$ tunes the nonlinearity. The output voltage $V_{\text{out}}$ is measured with a 8-GHz-analog-bandwidth 40-GS/s oscilloscope (Agilent DSO80804B). Feedback passes through a cavity with a dielectric object (2 cm $\times$ 4 cm water-filled container) that is positioned in 2-D using Thorlabs (LTS150) and Zaber Technologies (TLSR150B) translations stages. (b) Cavity pulse-response. We inject a 0.1 ns EM pulse of amplitude 1.5 V through the TX antenna and measure at the RX antenna. The time at which the radiation arrives at RX is a measure of the path length of EM energy through the cavity.}
\end{figure}

In our experimental system, the NLE is an input/output circuit based on the design from Ref.  \cite{Illing2006}. We use an aluminum two-dimensional (2-D) quarter-stadium-shaped RF cavity for a wave-chaotic scattering scene \cite{Alt1995}. As shown in Fig.\ \ref{fig:Figure1}a, the EM field emanating from the cavity is fed into a nonlinear circuit through a broadband (20 MHz - 2 GHz) receiving antenna (RX). The output of the circuit is fed back into the cavity through an identical transmitting antenna (TX), creating a closed feedback loop. Inside the cavity is a subwavelength dielectric object. 

The complex field inside of the cavity interacts multiple times with the object due to reflections from the cavity's walls (illustrated by a complex ray path in Fig.\ \ref{fig:Figure1}a) and from many passes of the RF signal through the nonlinear feedback loop. Due to these multiple interactions, small object movements change the structure of the field. These changes alter the dynamical state of the system. The output of the nonlinear circuit is filtered such that its maximum frequency is 2 GHz, and thus the RF signal has $\lambda \ge$ 15 cm. 

The NLE in the EM feedback loop induces quasi-periodic oscillations with multiple incommensurate frequencies in the output voltage $V_{\text{out}}$ of the nonlinear circuit. As the object moves inside the cavity, the frequencies of the quasi-periodic oscillations shift independently and provide a unique fingerprint of the object's location in 2-D. Thus, we map the position of the object in both the $x$ and $y$ directions by monitoring changes of a single scalar voltage $V_{\text{out}}$. 

Before describing our results, we first characterize our wave-chaotic cavity using its pulse response. Shown in Fig.\ \ref{fig:Figure1}b, our cavity produces a complicated pulse response (typical of wave-chaotic systems) with a quality factor $Q$ = 174 at a frequency of 1.77 GHz (the most prominent frequency in the quasi-periodic oscillations). As a result, broadcasting a continuous-wave signal into this cavity forms a complex interference pattern for each contained frequency (generic cavities tend to display such wave chaos; only cavities with a high degree of symmetry display simple interference patterns) \cite{Stockman1990}.

The pulse response of a wave-chaotic environment has been exploited to sense the appearance of an object in a scattering medium \cite{Taddese2010} or the location of a perturbation on the surface of a scattering medium \cite{Ing2005}. These techniques rely on measuring changes to the pulse response and have demonstrated a spatial sensitivity of $\sim\lambda$. Our own work is inspired by these achievements, where we use a continuous-time nonlinear feedback loop to achieve deep subwavelength position resolution.

Conventional oscillators using time-delayed nonlinear feedback use a nonlinear element whose output is  amplified and coupled back to the input through a single feedback loop that delays the signal by a fixed amount. These systems can display a variety of behaviors including periodic oscillations, quasi-periodicity, and chaos. Oscillators using time-delayed feedback have been designed using high-speed commercial electronics or lasers to generate complex signals with frequency bandwidths that stretch across several gigahertz \cite{Kouomo2005,Zhang2009,Murphy2010}.

Thus, the system shown in Fig.\ \ref{fig:Figure1}a combines the sensitivity of a dynamical state from a high-speed nonlinear-feedback oscillator with the sensitivity of the EM field in a wave-chaotic cavity. The time delays of the feedback in this system are the propagation times for the EM energy to transmit through the cavity, rather than a single time-delay. The values of the delays and their respective gains form a continuous delay distribution (proportional to the cavity's pulse response) that is uniquely defined for each position of the enclosed object. Due to the nonlinear feedback, the system's dynamics are highly sensitive to changes in this distribution of delays. Measuring the scalar variable $V_{\text{out}}$, we monitor dynamical changes in the system and sense the object's movements.

We first demonstrate this idea qualitatively along a one-dimensional (1-D) object path. We fix $V_{\text{bias}}$ in the NLE to exhibit periodicity at $x =$ 0 mm and measure the time evolution in $V_{\text{out}}$ for object positions $x =$ 0 mm -- 12 mm in 10 $\mu$m steps. The system changes between periodicity (P) from $x =$ 0 mm -- 1.4 mm, quasi-periodicity (QP) from $x =$ 1.4 mm -- 8 mm, and two different time-evolving chaotic states (C1 and C2) from $x =$ 8 mm -- 9.8 mm and $x =$ 9.8 mm -- 12 mm, respectively. Chaotic state C1 contains chaotic-like breathers and C2 exhibits a relatively flat bandwidth from 20 MHz -- 2 GHz.

\begin{figure}[tbp]
\begin{center}
 \resizebox{7.5cm}{!}{\includegraphics{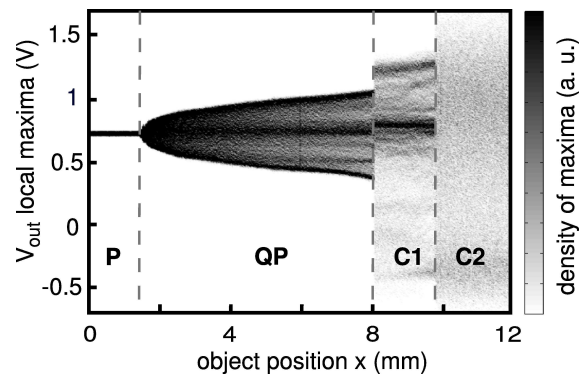}}
\end{center}
\caption{\label{fig:Figure2} Bifurcation Diagram. We store and analyze time series of $V_{\text{out}}$ at each object position along a 1-D path in the $x$ direction. The local maxima of each time series is plotted as a function of object position $x$.}
\end{figure}

The observed dynamical changes fall into one of two categories: an abrupt change in the dynamical state (known as a bifurcation) or small shifts in the frequency components and amplitudes of $V_{\text{out}}$. A bifurcation diagram illustrates the qualitative dynamical changes in Fig.\ \ref{fig:Figure2}. Our results show  dynamical changes from subwavelength movements of a subwavelength object.

\begin{figure}[tbp]
\begin{center}
 \resizebox{\figurewidth}{!}{\includegraphics{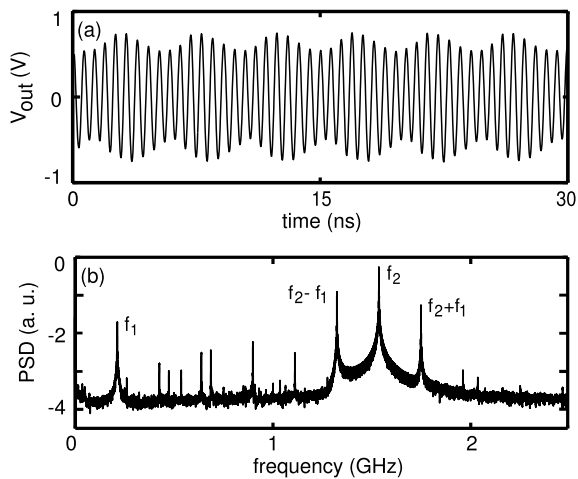}}
\end{center}
\caption{\label{fig:Figure3}(a) Temporal evolution of a typical quasi-periodic state used for position sensing.  (b) Logarithmic power spectral density (PSD) of this signal.}
\end{figure}

To go beyond the qualitative detection of movement, we tune $V_{\text{bias}}$ so that $V_{\text{out}}$ is in a quasi-periodic state (QP in Fig.\ \ref{fig:Figure2}) for all object positions of interest. The incommensurate frequencies of a QP state are not phase-locked and hence can shift independently with respect to object translations. In addition, incommensurate frequencies help eliminate interference nodes (blind spots) of the illuminating EM waves in the cavity, where each frequency has a complex interference pattern that covers the blind spots of another. An example time series and frequency spectra for a fixed object position are seen in Fig.\ \ref{fig:Figure3}a and Fig.\ \ref{fig:Figure3}b, respectively. Recall that, though the system's dynamics are QP in time, the EM energy inside of the cavity is chaotic.  

Tracking the object entails measuring shifts in the QP frequency components. In Fig.\ \ref{fig:Figure3}b, we highlight two peaks in the spectrum at frequencies denoted by $f_1$ and $f_2$. The frequency harmonics at ($f_2 - f_1$) and ($f_2 + f_1$) are used to improve the signal-to-noise ratio (SNR) of these frequencies. Averaging independent measures of $f_1$ and $f_2$ reduces statistical errors and increases their SNR. To follow changes in the frequencies of $V_{\text{out}}$ with high precision, we use a nonlinear least-squares-fit to a model for a four-tone QP signal \cite{Weaver2006}, resulting in a 2.4 kHz frequency resolution (approximately 0.5$\%$ of the total observed experimental frequency shifts). 

\begin{figure}[htbp]
\begin{center}
 \resizebox{8.0cm}{!}{\includegraphics{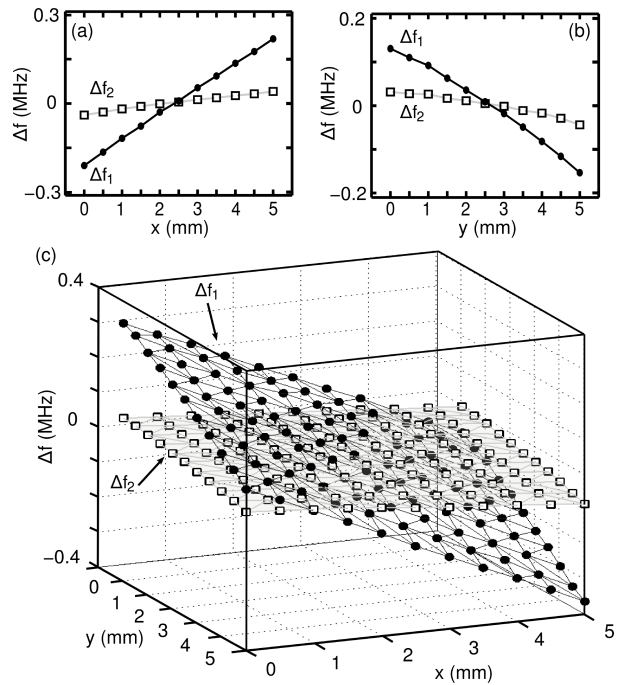}}
\end{center}
\caption{\label{fig:Figure4} Frequency shifts $\Delta f_1$ and $\Delta f_2$ of the QP state as the object translates along the (a) $x$ direction, where the fit from Eq.\ (\ref{eq:oneD1}) yields $a_1 =$ 2.8 mm/kHz, $a_2 =$ -8.7 mm/kHz, $c_0$ = 27,113.0 mm, $c_1$ = -439.2 and $c_2$ = 1.5 mm$^{-1}$, and (b) along the $y$ direction, where the fit from Eq.\ (\ref{eq:oneD2}) yields $b_1 =$ 0.8 mm/kHz, $b_2 =$ -0.5 mm/kHz, $d_0$ = -73,452.0 mm, $d_1$ = 242.0, and $d_2$ = -2.0 mm$^{-1}$. (c) Frequency shifts $\Delta f_1(x,y)$ and $\Delta f_2(x,y)$ for object translations $(x,y)$ in a 5 mm $\times$ 5 mm grid of positions. The planar fits from Eqs.\ (\ref{eq:twoD1}) and \ (\ref{eq:twoD2}) yield $\alpha_1$ = -84.68 kHz/mm, $\alpha_2$ = -15.20 kHz/mm, $\beta_1$ = -56.74 kHz/mm, $\beta_2$ = -14.75 kHz/mm, $\epsilon_1$ = 11.72 MHz and $\epsilon_2$ = 2.43 MHz. The measured determinant $|\alpha_1 \beta_2 - \alpha_2 \beta_1|=$ 386 kHz$^2$/mm$^2$ with an error of 5.8 kHz$^2$/mm$^2$.}
\end{figure}

To demonstrate 1-D position sensing, we translate the object along the path $x$ = 0 mm - 5 mm while $y$ = 2.5 mm. We then translate the object along an orthogonal path $y$ = 0 - 5 mm while $x$ = 2.5 mm. Shown in Fig.\ \ref{fig:Figure4}a and Fig.\ \ref{fig:Figure4}b, the measured frequency shifts $\Delta f_1$ and $\Delta f_2$ are plotted for 1-D paths along the orthogonal $x$ and $y$ directions, respectively. We separately fit $\Delta f_1$ and $\Delta f_2$ in the $x$ and $y$ directions with second order polynomials
\begin{equation}
a_1 \Delta f_1(x) + a_2 \Delta f_2(x) = c_0 + c_1 x + c_2 x^2,
\label{eq:oneD1}
\end{equation}
\begin{equation}
b_1 \Delta f_1(y) + b_2 \Delta f_2(y) = d_0 + d_1 y + d_2 y^2.
\label{eq:oneD2}
\end{equation}
We optimize the coefficients $a_i$ and $c_i$ ($b_i$ and $d_i$) using a nonlinear least-squares-fit to a model for the object position. The root-mean-square (RMS) errors for the frequency shift map is 1.45 kHz (0.86 kHz) along $x$ $(y)$. By inverting these maps, we calculate the measured object positions. The RMS error between the actual and measured positions is 9.2 $\mu$m (23.7 $\mu$m) for $x$ $(y)$, which demonstrates a resolution of $\sim\lambda/10,000$ along orthogonal 1-D directions (recall $\lambda \ge$ 15 cm).

Tracking the object's position in both the $x$ and $y$ directions simultaneously requires two independently changing observables. In our system, we observe a single scalar variable $V_{\text{out}}$ that oscillates with primary  frequencies $f_1$ and $f_2$. We fit the frequency shifts $\Delta f_1(x,y)$ and $\Delta f_2(x,y)$ for object positions $(x,y)$ in a 5 mm $\times$ 5 mm area (Fig.\ \ref{fig:Figure4}c) and approximate them as planes
\begin{equation}
\Delta f_1(x,y) = \alpha_1 x + \beta_1 y + \epsilon_1, 
\label{eq:twoD1}
\end{equation}
\begin{equation}
\Delta f_2(x,y)  = \alpha_2 x + \beta_2 y + \epsilon_2. 
\label{eq:twoD2}
\end{equation}
Using Cramer's rule, we show that $|\alpha_1 \beta_2 - \alpha_2 \beta_1| \ne 0$ to verify the planes are linearly independent in this area and allow us to simultaneously measure $x$ and $y$ coordinates. 

In the 1-D case, we have the freedom to optimize the fitting parameters in Eqs.\ (\ref{eq:oneD1}) and \ (\ref{eq:oneD2}) for $x$ and $y$ separately. In the 2-D case, all of the fitting parameters $\alpha_i$, $\beta_i$, and $\epsilon_i$ in Eq.\ (\ref{eq:twoD1}) and Eq.\ (\ref{eq:twoD2}) are present in the solutions for both $x$ and $y$. Thus, we cannot optimize the fits in the $x$ and $y$ directions separately and instead use the fitted planes for our frequency maps. 

This constraint, combined with the approximation that these surfaces are planar, limits our 2-D resolution. A planar fit of $\Delta f_1(x,y)$  $(\Delta f_2(x,y))$ gives a RMS frequency error of 4.17 kHz (7.26 kHz) and a RMS position error of 370 $\mu$m (650 $\mu$m) for $x$ $(y)$, yielding a 2-D resolution of $\sim\lambda/300$. Higher order fits do not improve the resolution due to noise in our measurements. This 2-D frequency-mapping serves as the calibration for objects of this shape and must be reacquired for different shaped objects. 

For comparison, a scanning near-field microwave microscope uses RF frequency shifts to achieve subwavelength sensitivity ($\sim\lambda/$750,000) of near planar surfaces \cite{Azar1999}. In contrast, our system uses nonlinear feedback to internally generate multiple independent frequencies and measures multiple degrees-of-freedom using a single scalar variable. Moreover, it uses a stationary pair of antennas to extract 2-D spatial information of a 3-D object, making it free of mechanically-moving parts.

We conjecture that our method can be implemented using EM waves in the visible part of the spectrum. Semiconductor lasers with time-delayed optical feedback are known to display complex dynamical behaviors in which the output intensity varies in time, including quasi-periodicity \cite{Ikeda1980,Mork1990,Fischer1974}. Furthermore, optical wave chaos has been demonstrated using optical cavities \cite{Nockel1997,Gmachl1998,Gensty12005}. We envision a completely optical version of our technique where a laser receives feedback from a wave-chaotic optical cavity. Such a system will be capable of tracking an object on a sub-nanometer scale.

Understanding the full potential of this method will require studies in both wave chaos and nonlinear dynamics. Our results suggest that one can position sense in 3-D using a QP state with three independent frequencies. This type of dynamical state is possible in our system but requires further study to create a QP state for a 3-D volume of interest. More independent observables could also be introduced into the system using two or more feedback loops external to the cavity, where each loop is independently band-limited to prevent cross talk.

In the future, we see several options to improve the system's resolution. Increasing the number of frequency harmonics through nonlinear mixing gives additional measures of the independent modes and improves the system's SNR. Also, the cavity $Q$ is proportional to the number of interactions between the subwavelength object and the EM energy inside of the cavity, and thus the resolution of this technique should also scale with $Q$.

We believe that our system will have applications beyond position sensing. Subwavelength scatterers are often treated as point-like objects; our approach is sensitive to the shape and orientation of the subwavelength scatterer. Also, similar to \cite{Lobkis2009}, analyzing dynamical states can monitor changes in the EM properties of materials in the cavity.

To the best of our knowledge, our approach is the first to measure multiple spatial degrees-of-freedom on a subwavelength scale using a single scalar signal. Using a QP analog of the Larsen effect, we combine a nonlinear feedback oscillator with multiple EM reflections in a scattering environment to exploit the inherent sensitivity of wave chaos, adding an alternative to the short list of super-resolution techniques.

We gratefully acknowledge Zheng Gao with help in designing NLE and the financial support of the U.S. Office of Naval Research grant $\#$ N000014-07-0734.

\bibliographystyle{apsrev}
\bibliography{references_subwavelength}
\end{document}